\documentclass[prl,twocolumn,floatfix,showpacs,superscriptaddress]{revtex4}
\usepackage{graphicx}
\usepackage{amsmath}
\begin{document}

\title{Kondo effect in transport through molecules adsorbed on metal surfaces: \\
from Fano dips to Kondo peaks}
\author{J.M. Aguiar-Hualde}
\affiliation{Departments de F\'{\i}sice J.J. Gambians, Faculae de
Ciencias Exactas, Universidad de Buenos Aires, Ciudad Universitaria,
1428 Buenos Aires, Argentina.}
\author{G. Chiappe}
\affiliation{Departments de F\'{\i}sice J.J. Gambians, Faculae de
Ciencias Exactas, Universidad de Buenos Aires, Ciudad Universitaria,
1428 Buenos Aires, Argentina.} \affiliation{Departamento de
F\'{\i}sica Aplicada, Unidad Asociada del Consejo Superior de
Investigaciones Cient\'{\i}ficas and Instituto Universitario de
Materiales, Universidad de Alicante, San Vicente del Raspeig,
Alicante 03690, Spain.}
\author{E. Louis}
\affiliation{Departamento de F\'{\i}sica Aplicada, Unidad Asociada del
Consejo Superior de Investigaciones Cient\'{\i}ficas and
Instituto Universitario de Materiales, Universidad
de Alicante, San Vicente del Raspeig, Alicante 03690, Spain.}
\author{E.V. Anda}
\affiliation{Departamento de  F\'{\i}sica, Pontificia Universidade
Cat\'olica do Rio de Janeiro (PUC-Rio), 22452-970, Caixa Postal:
38071 Rio de Janeiro, Brazil.}
\date{\today}
\begin{abstract}

The Kondo effect observed in recent STM experiments on transport
through CoPc and TBrPP-Co molecules adsorbed on  Au(111) and Cu(111)
surfaces, respectively, is discussed within the framework of a
simple model (Phys. Rev. Lett. {\bf 97}, 076806 (2006)). It is shown
that, in the Kondo regime and by varying the adequate model
parameters, it is possible to produce a crossover from a conductance
Kondo peak (CoPc) to a conductance Fano dip (TBrPP-Co). In the case
of TBrPP-Co/Cu(111) we show that the model reproduces the
changes in the shape of the Fano dip, the raising of the Kondo
temperature and shifting to higher energies of the dip minimum when
the number of nearest neighbors molecules is lowered. These
features are in line with experimental observations indicating that
our simple model contains the essential physics underlying the
transport properties of such complex molecules.

\end{abstract}
\maketitle

Since the work of Madhavan {\it et al} \cite{MC98} on the Kondo
effect \cite{Ko64,He97} in transport through a Co atom adsorbed on
an Au(111) surface, a great deal of attention has been devoted to
investigate, both theoretical and experimentally, such effect in
either isolated Co atoms \cite{UK00,NK07} or in Co-containing
molecules \cite{ZL05,Cr05,CL06,ID06} adsorbed on metal surfaces. The
possibility of tuning the Kondo temperature \cite{ZL05,ID06} has
been recently demonstrated, increasing considerably the interest of
these systems. In particular it has been shown that, the
characteristics, and even the existence, of the Kondo resonance can
be controlled by distorting a CoPc molecule adsorbed on a Au(111)
surface \cite{ZL05}. More recently \cite{ID06} the Kondo temperature
in a TBrPP-Co molecule adsorbed on a Cu(111) surface has been
increased by decreasing the number of nearest-neighbor TBrPP-Co
molecules. The authors of \cite{ID06} argued that this is due to a
reduction of surfaces states when the number of nearest neighbors
molecules around a given one is increased.

One of the most interesting aspects of those two works is that while
in the case of CoPc/Au(111) a Kondo peak was observed, in the
experiments on TBrPP-Co/Cu(111) the conductance showed a Fano dip,
as in isolated Co atoms adsorbed on metal surfaces
\cite{MC98,UK00,NK07}. Those two molecules, although largely
different, have two outstanding similarities: i) both have the Co
atom in the their geometric centers, and, ii) their STM images show
four clearly defined lobes \cite{ZL05}. The enormous complexity of
the electronic structure of these molecules and the many-body
physics involved hinder a detailed account of these
systems. One of the main goals of the present work is to show, using
a simple model that catches the most prominent features of
both molecules, under which circumstances dips or peaks appear in
the transmission.

A model Hamiltonian  was taken assuming the following small
atomic arrangement \cite{CL06}: a central site with a single atomic
orbital, a strong Coulomb repulsion at the Co atom and four lobes of
the molecule described by four atomic orbitals placed on a square,
which center is the Co atom (see  Fig. 1 of Ref. \cite{CL06}). Two
additional orbitals located above and below the Co atom are included
to represent the apex of the STM tip and an atom on the metal
surface, respectively.

The Hamiltonian investigated here takes  the form,
\begin{equation}
{\hat H} = \sum_{i\sigma}\epsilon_i
c^{\dagger}_{i\sigma}c_{i\sigma} +
\sum_{<ij>;\sigma}t_{i,j}c^{\dagger}_{i\sigma}
c_{j\sigma}+
Un_{Co\uparrow}n_{Co\downarrow}
\label{eq:H}
\end{equation}
where $c^{\dagger}_{i\sigma}$ creates an electron at site
$i$ and spin $\sigma$ and $n_{Co\sigma}$ is the occupation operator
associated to Co. The cluster is connected to the STM tip and the
metal surface both represented by energy independent self-energies.
The parameter $t_{i,j}$ is the hopping between atomic orbitals
located on sites $i$ and $j$ (the symbol $<>$ in Eq. (1) indicates
that $i\neq j$), each orbital has an energy $\epsilon_i$, and the
local Coulomb repulsion on Co is described by $U$.

The hopping matrix elements incorporated in Ref. \cite{CL06}
were: $t_{Co,t}$ (Co/STM tip), $t_{Co,m}$ (Co/metal surface),
$t_{Co,l}$ (Co/molecule lobes) $t_{l,l}$ (inter-lobe hopping) and
$t_{l,m}$ (lobes/metal surface). In this work we incorporate an
additional parameter: the direct hopping between the tip and the
lobes $t_{t,l}$. The reason to do so is that Fano antiresonances
appear whenever the main current flows through sites that have
laterally attached strongly correlated sites \cite{UK00,MB06}. On the
other hand, this is justified by the fact that, in TBrPP-Co/Cu(111),
the $d_{z^{2}}$ orbital (the one that is likely responsible for the
Kondo effect, see below) is much deeper than in CoPc/Au(111) (-0.7
and -0.15 eV, respectively) indicating that,  in TBrPP-Co, $t_{t,l}$
and $t_{Co,t}$ may  at least be comparable. One lead is attached to
Co and describes the STM tip. The other lead (the metal surface) is
attached to the surface atom below Co
 and to the lobes.

When the cluster is connected to electrodes, the transmission across the system
is given by  $T(E)=\frac{2e^2}{h}{\rm Tr}[t^{\dagger}t]$ \cite{La57},
and the conductance is ${\mathcal G}=T(E_F)$, where $E_F$ is
the Fermi level. In this expression,  matrix $t$ is
$t= \Gamma_{\rm U}^{1/2}G^{(+)}\Gamma_{\rm L}^{1/2}=
\left[\Gamma_{\rm L}^{1/2}G^{(-)} \Gamma_{\rm U}^{1/2} \right]^{\dagger}$,
where $\Gamma_{\rm U(L)}= i(\Sigma^{(-)}_{\rm U(L)}-\Sigma^{(+)}_{\rm U(L)})$,
$\Sigma^{(\pm)}_{\rm U(L)}$  being the
self-energies of the upper (U) and lower (L) leads, STM tip and gold surface,
respectively.  Superscripts (+) and (-)
stand for retarded and advanced.
The Green function  is written as \cite{MW92},
$G^{(\pm)}=\left(\left[G_0^{(\pm)}\right]^{-1}-
\left[\Sigma_U^{(\pm)}+\Sigma_L^{(\pm)}\right]\right)^{-1}$,
where $G_0^{(\pm)}$ is the Green function  of the isolated  cluster.
In this work  $G_0^{(\pm)}$ is obtained either by means of  a finite
$U$ slave bosons (SB) approach \cite{KR86,VS06,AC06,ACun} or by exact
diagonalization in which case the result corresponds to the
Embedding Cluster Approximation (ECA) method
\cite{CL06,MB06,FC99,CL05}.

In carrying out calculations,  we have taken the self-energies
attached to the tip and metal site below Co $\Sigma_U^{(\pm)}=\mp 0.2 i$ eV, whereas
that attached to the lobes $\Sigma_L^{(\pm)}=\mp t_{l,m}^2\rho_{m} i$, where
$\rho_{m}$ is the density of states at the metal surface that will be varied to simulate
the presence or absence of surface states. In addition we assume half-filling (one
electron per site) and take  $\epsilon_{Co}=-U/2$ and the rest of atomic
orbitals lying at zero energy (hereafter  the Fermi level $E_{F}$
of the whole system will be taken as the zero of energies).

Other model parameters have been varied aiming to identify their
role in the behavior of this system, but always over ranges such
that the main features in the conductance occur in energy scales
similar to the experimental ones \cite{ID06}. It is worth mentioning
that in the present work $U$ was taken always smaller than 3 eV,
while in \cite{CL06} a value of 8 eV was  assumed. We note that the
latter value is probably too large, as indicated by a recent study
of atomic Co on a Cu(100) surface \cite{NK07}. Although screening in
the Co/Cu(100) is probably more important than in the present case,
we adopt the more conservative lower values. All calculations were
done at zero temperature.

\begin{figure}
\includegraphics[width=3.4in,height=3.4in]{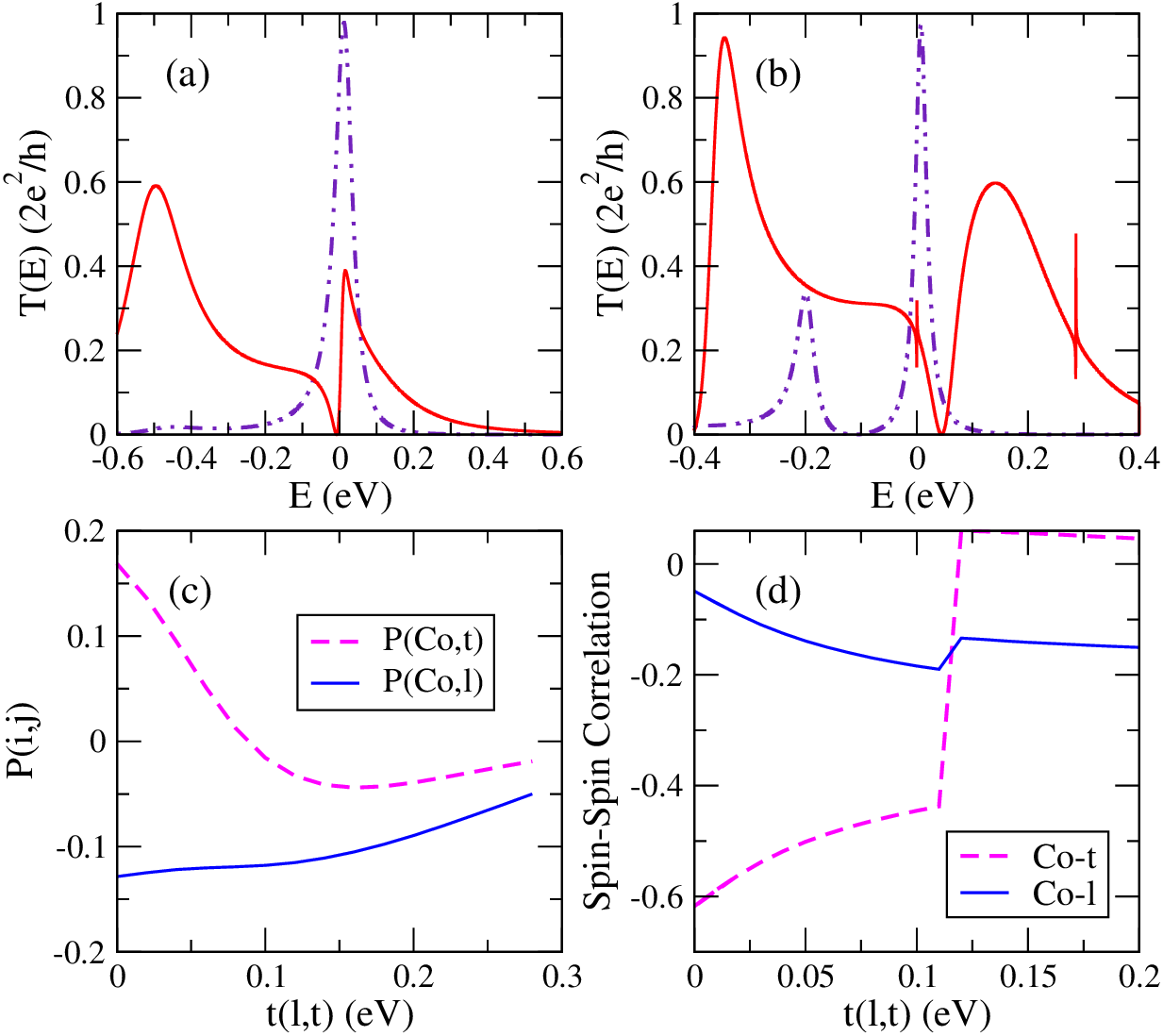}
\caption{(Color on line). Upper panels: transmission {\it T(E)} (in
units of the conductance quantum) versus the energy {\it E} (in eV)
referred to the Fermi energy, calculated by means of the SB (a) and
the ECA (b) methods (see text); results for  $t_{l,t}=0$ (green
chain line)  $t_{l,t}=0.08$ eV (continuous red line) are shown. (c)
Non-diagonal elements of the density matrix versus $t_{l,t}$ as
calculated by means of the SB approach. (d) Spin-spin correlation
versus $t_{l,t}$ calculated by means of the ECA method. The rest of
the parameters used in the calculations are: $t_{Co,t}=0.08 eV$,
$t_{Co,l}=-0.14 eV$, $t_{Co,m}=0$, $t_{l,l}=-0.2 eV$, $t_{l,m}=0.14
eV$, $U$ =1.6 eV and $\rho_{m}=5 eV^{-1}$.}

\end{figure}

The results depicted in Fig. 1  illustrate how switching on the
lobe/tip hopping produces a crossover from a Kondo peak to a Fano
dip. In order to make more apparent the crossover we took
$t_{Co,m}$=0. The crossover is also clearly noted in the numerical
results for the non-diagonal elements of the density matrix shown in
Fig. 1c. The latter is calculated using the standard expression,
$P(i,j)=\frac{1}{2\pi}\int_{-\infty}^{\infty}G^<(i,j;E)dE$,
where $G^<(i,j;E)$ is the lesser Green function of the whole system.
It is noted that $P(Co,t)$ is strongly reduced and tends to
zero as the  lobes/tip hopping becomes $t_{l,t} > t_{Co,t}$, triggering 
the crossover in the transmission from a Kondo peak to a Fano dip.
This indicates that, in the latter situation, the current flowing
through the Co reduces significantly. There is a concomitant small
reduction of $P(Co,l)$ that however, remains significantly
greater (in absolute value) than $P(Co,t)$ for
$t_{l,t} > t_{Co,t}$, as shown in the figure. The drastic reduction of
the tip/Co current permits the appearance of a destructive
interference among the electrons going directly from the tip to the
metal through the lobes and those performing the same trajectory but
visiting the Co atom. This last path is possible due to the Kondo
resonance located at the Co. From this point of view the Co atom
acts as a correlation impurity laterally coupled to the conduction
channel. On the other hand, the results for the spin-spin
correlation calculated with the ECA method shown in Fig. 1d clearly
indicate that the strong anti-ferromagnetic correlation between Co
and the tip characteristic of the Kondo regime for
$t_{Co,t}>t_{t,l}$ \cite{CL06} is destroyed as $t_{t,l}$ increases.
 In this case the Kondo regime is driven by the
Co/lobes correlation that remains always
anti-ferromagnetic. It is worth noting that the spin-spin
correlation shown in the figure changes abruptly because it corresponds
to the isolated cluster which is exactly diagonalized when the ECA method is
used.
\begin{figure}
\includegraphics[width=3.1in,height=3.8in]{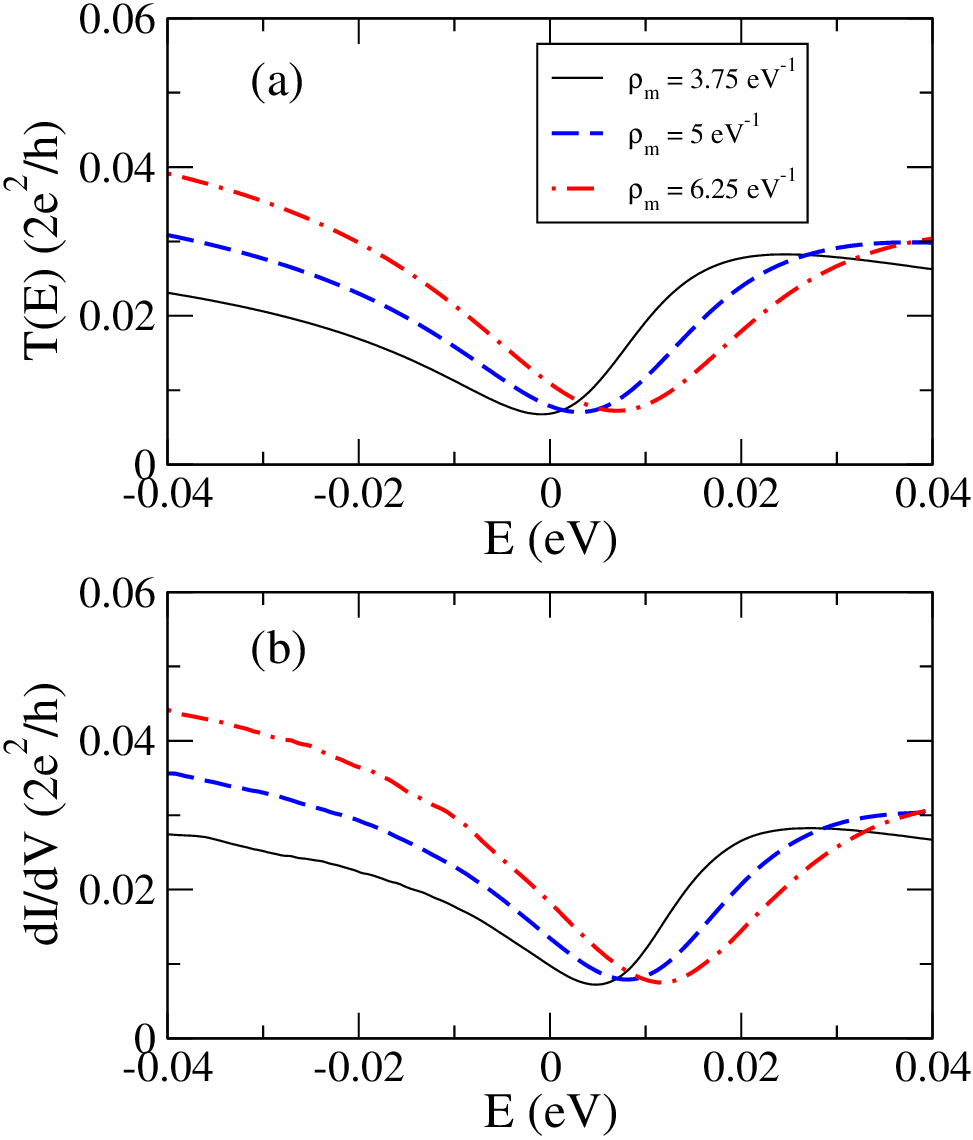}
\caption{(Color on line). Transmission $ T(E)$ (a) and  differential conductance
$dI/dV$ (b) (both in units of the conductance quantum) versus
either the energy {\it E} referred to the Fermi energy or the bias potential
$V_{bias}$
(in eV) as calculated
by means of the SB method. Results for three values of the metal density of states are
shown. The rest of the model parameters used in the calculations are:
$t_{Co,m}=0.04$ eV, $t_{Co,l}=-0.14$ eV, $t_{Co,t}=0.04$ eV, $t_{l,l} = -0.2$ eV,
$t_{l,m}=0.14$ eV,  $t_{l,t}=0.032$ eV and $U=1.6$ eV.}
\end{figure}

Results obtained by means of either the SB or ECA methods (Fig.
1a,1b) are very similar, they only differ at a quantitative level.
The noticeable differences between the two approaches make this
agreement particularly relevant. On the other hand, although both
methods are reliable around the Fermi level, only  the SB approach
allows an easy implementation of Keldysh formalism \cite{MW92} to
treat the out-of-equilibrium problem, as discussed below. Taking
note of the latter, and recalling that the SB method is numerically
faster, the results presented and discussed hereafter are all
obtained with that method. In the case of the Fano dip, we note the
presence of a broad peak at around -0.5 eV. This peak almost
coincides with that observed experimentally at -0.7 eV that was
ascribed to the $d_{z^2}$ orbital, the one probably responsible of
the strong correlation effects discussed here (see also below). It
should be note that, the latter peak shows up below $E_{F}$ if both
$t_{l,l}$ and $t_{Co,l}$ are negative. When they are assumed to be
positive the results change to $T(-E)$ and if only one of them is
negative the transmission is modified resembling less the
experimental observations (a wider dip and a narrower peak below
$E_{F}$).

As remarked above, it was suggested in \cite{ID06} that  there is a
correlation between the number of nearest neighbors nn around a
molecule and the density of states $\rho_{m}$ seen by the peripheral
atoms of the molecule (the larger nn the smaller $\rho_{m}$). The
effects of varying the density of states at the metal surface on the
Fano dip are illustrated in Fig. 2. In these calculations we have
taken a value of $t_{Co,m}$ similar to that of $t_{Co,t}$ as it
seems more realistic. Results for the transmission versus energy are
shown in the upper panel, whereas the differential conductance
versus bias voltage is shown in the lower panel. They were 
calculated by implementing Keldysh formalism into the SB approach
along the lines discussed in Ref. \cite{MW93}. It is noted that the
two results are remarkably similar, as expected whenever the bias
voltage is not very large \cite{LV03}. This supports the
validity of the comparison of the transmission $T(E)$ with the
experimental results associating the external applied field to the
variable $E$. In figure 2  we note that, increasing the density of
states, the Fano antiresonance becomes wider and dipper and shifts
to higher voltages. The first effect implies an increase of the
Kondo temperature, in agreement with the proposal made in Ref.
\cite{ID06}. Concerning the shift, it was mentioned in \cite{ID06}
that the voltage at which the minima in the Fano dips occurred,
changed when the number of neighboring molecules was varied,
although unfortunately a well-defined tendency was not mentioned. We
also note changes in the shape of the dip (particularly in the
relative height of the left and right shoulders) also observed in
the experimental study of Ref. \cite{ID06}.

Fig. 3a further illustrates the crossover from Fano dip to Kondo
peak investigated in this work. The figure depicts results for the
transmission versus energy for three values of the lobes/tip
hopping, namely, $t_{l,t}$= -0.18 eV, 0.0  and 0.06 eV. The
transmission pattern changes from Fano dip-like to Kondo peak-like
and back to Fano dip-like. This behavior resembles the results
reported in \cite{UK00} which showed a similar oscillation of the
tunneling DOS as the distance from the tip to the impurity (a Co
atom adsorbed on a metal surface) was varied. Increasing the Coulomb
repulsion parameter $U$ (see Fig. 3c) sharpens both the Fano dip and
the Kondo peak, although scarcely affects the structure away the
Fermi level that shows up in the first case. The sharpening
of the structures results from the reduction of the Kondo
temperature when U is increased. It is interesting to note that the
Fano anti-resonance coexists with the broad peak below the Fermi
level, while the Kondo peak does not. This is further illustrated
by the transmission results shown in Fig. 3b for two values of the
lobe/lobe hopping. The results are conclusive: the well defined
Kondo peak that shows up for the larger value of $t_{l,l}$,
transforms into a much broader feature peaked at around -0.15 eV
when that hopping is reduced. Note that the latter feature may
actually be that reported in \cite{ZL05} for the undistorted
molecule which, as pointed out before, does not show a Kondo peak.
The effects of increasing $U$ (see Fig. 3d) are similar to those noted in the case of Fig 3c.

 \begin{figure}
\includegraphics[width=3.4in,height=3.4in]{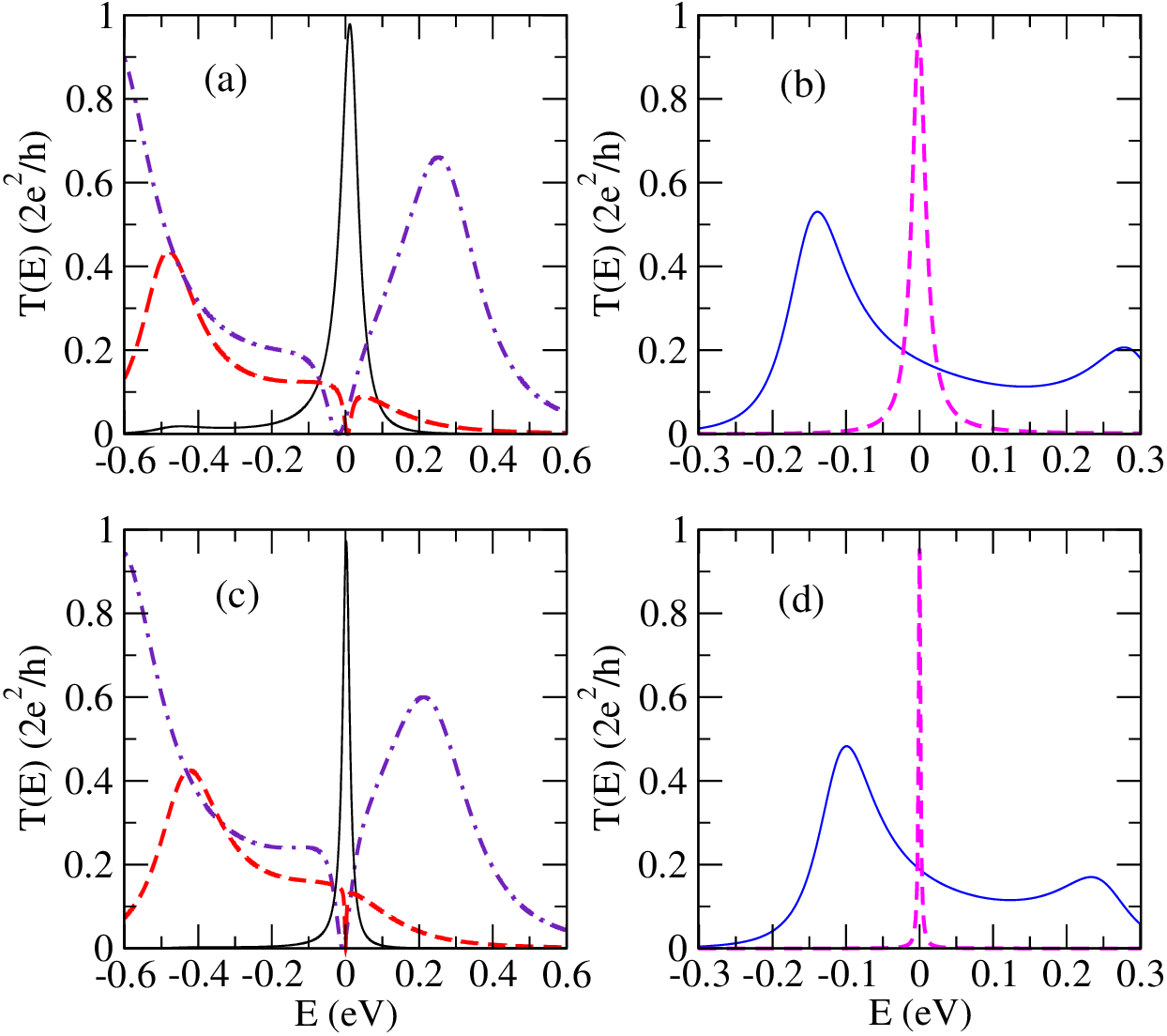}
\caption{(Color on line). Transmission {\it T(E)} (in units of the
conductance quantum) versus the energy {\it E} (in eV) referred to
the Fermi energy, calculated by means of the SB  method (see text).
The parameters used in the calculations are hereafter given:
$t_{Co,m}=0.0$ eV, $t_{Co,l}=-0.14$ eV, $t_{Co,t}=0.08$ eV. Left
panels: $t_{l,l} = -0.2$ eV, $t_{l,m}=0.14$ eV,  $t_{l,t}$=-0.18 eV
(green chain line) 0 eV (continuous black line) and 0.06 eV (broken
red line) and $U=1.6$ eV (a) and $U=2.3$ eV (c). Right panels:
$t_{l,l}$ = 0.05 eV (continuous blue line) and 0.3 eV (broken
magenta line), $t_{l,m}=0.14$ eV,  $t_{l,t}=0.0$ eV, and $U=1.6$ eV
(b) and $U=2.3$ eV (d).}
\end{figure}

Summarizing, with the help of a simple model we have been able to
discuss the main features of the Kondo regime observed either as a
Fano dip in TBrPP-Co/Cu(111) or as a Kondo peak in CoPc/Au(111). We
showed that, by varying some of the model parameters, it was
possible to produce a crossover in the conductance from a Fano dip
to a Kondo peak. The model not only allows to discuss the physics
around the Fermi level, but in the Kondo regime also accounts for
features in the experimental conductance that coexist with the Fano
dip, or does not exist when the transmission has a peak. In
addition, we have been able to confirm the role of the surface
density of states in defining the Kondo temperature in the
TBrPP-Co/Cu(111), as suggested by the authors of that study. Our
analysis illustrates the usefulness of simple models in which strong
correlation effects preclude a full {\it ab initio} study.
\acknowledgments
Financial support by the spanish MCYT (grants FIS200402356, MAT2005-07369-C03-01
and NAN2004-09183-C10-08), the Universidad de Alicante, the Generalitat Valenciana
(grant GRUPOS03/092 and grant GV05/152), the Universidad de Buenos
Aires (grant UBACYT x115) the argentinian CONICET and the Brazilian agencies CNPq and Faperj
are acknowledge. GC is thankful to the spanish "Ministerio de Educaci\'on y Ciencia" for a Ram\'on
y Cajal grant.

\end{document}